\documentclass[11pt,twocolumn]{article}

\usepackage{amsmath, amsthm, amsfonts}
\usepackage{setspace}
\usepackage{authblk}
\usepackage[pdfborder={0 0 0}]{hyperref}
\usepackage[utf8]{inputenc}
\usepackage[english]{babel}			
\usepackage{color}							
\usepackage{exscale}						
\usepackage{times}							
\usepackage{amssymb}						
\usepackage{graphicx}						
\usepackage{pgf,pgfarrows,pgfnodes,pgfshade}	
\usepackage{float}
\usepackage{supertabular, booktabs}

\theoremstyle{definition}

\theoremstyle{remark}

\title{Novel model for wine fermentation including the yeast dying phase}
\author[1]{Alfio Borzì}
\affil{University of Würzburg, Institute for Mathematics\\
	D-97074 Würzburg\\
\url{alfio.borzi@mathematik.uni-wuerzburg.de}}
\author[2]{Juri Merger}
\affil{University of Würzburg, Institute for Mathematics\\
	D-97074 Würzburg\\
\url{juri.merger@mathematik.uni-wuerzburg.de}}
\author[3]{Jonas Müller}
\affil{Geisenheim University, Modeling and Systems Analysis \\
	D-65366 Geisenheim\\
\url{jonas.mueller@hs-gm.de}}
\author[4]{Achim Rosch}
\affil{Dienstleistungszentrum Ländlicher Raum (DLR) Mosel, Department of Winegrowing and Enology\\ D-54470 Bernkastel-Kues 
\url{achim.rosch@dlr.rlp.de}}
\author[5]{Christina Schenk*}
\affil{Trier University, Department of Mathematics\\
	D-54286 Trier\\
\url{christina.schenk@uni-trier.de}}
\author[6]{Dominik Schmidt}
\affil{Geisenheim University, Modeling and Systems Analysis \\
	D-65366 Geisenheim\\
\url{dominik.schmidt@hs-gm.de}}
\author[7]{Stephan Schmidt}
\affil{University of Würzburg, Institute for Mathematics\\
	D-97074 Würzburg\\
\url{stephan.schmidt@mathematik.uni-wuerzburg.de}}
\author[8]{Volker Schulz}
\affil{Trier University, Department of Mathematics \\
D-54286 Trier\\
\url{volker.schulz@uni-trier.de}}
\author[9]{Kai Velten}
\affil{Geisenheim University, Modeling and Systems Analysis \\
D-65366 Geisenheim\\
\url{kai.velten@hs-gm.de}}
\author[10]{Christian von Wallbrunn}
\affil{Geisenheim University, Microbiology and Biochemistry\\
D-65366 Geisenheim\\
\url{Christian.Wallbrunn@hs-gm.de}}
\author[11]{Michael Zänglein}
\affil{Bayerische Landesanstalt für Weinbau
und Gartenbau (LWG), Department of Winegrowing\\D-97209 Veitshöchheim\\
\url{michael.zaenglein@lwg.bayern.de}\newline Corresponding author:*}
\begin{document}
\maketitle
\pagenumbering{arabic}
\abstract{This paper presents a novel model for wine fermentation including a death phase for yeast and the influence of oxygen on the process. A model for the inclusion of the yeast dying phase is  derived and compared to a model taken from the literature. The modeling ability of the several models is analyzed by comparing their simulation results.}

\section{Introduction}
In food industry there is always a need to reduce production costs as, for example, the energy needed for cooling or heating to improve and 
maintain the quality of the product. For this purpose, mathematical simulation and optimization methods are investigated and applied 
to the production chain. This is the main objective of the project R\OE NOBIO, that focuses on the modeling and optimization of the 
wine fermentation process . Fundamental for this task is a model that is able to accurately describe the dynamics of the wine fermentation.\\
Wine fermentation is a very complex bio-chemical process that can be influenced in many ways by the environment and by the 
addition and subtraction of bio-chemical components. For example, oxygen can be added to the must as it plays a crucial role for yeast activity. This takes place in the phase where the yeast is unable to cope with the high concentrations of sugar and nutrients. But the yeast has to take up the added oxygen before it can react with the must as this leads to oxidation of the wine. More information on the effects of oxygen on fermentation can be found in Shea A.J. Comfort \cite{oxygenarticle}.\\
During fermentation yeast grows by metabolizing sugar in presence of nutrients such as assimilable nitrogen. The consumed sugar is converted into ethanol. However, ethanol inhibits yeast growth and a high amount of it causes cell death.\\
In the literature, many different mathematical systems that model the process of wine making based on different kinetics can be found. Usually the development of yeast, sugar, alcohol, and assimilable nitrogen concentration are taken into account. However, these models do not represent the behavior of yeast cells that is observed in real experiments. The growth phases of yeast can be found in Dittrich et al. \cite{mikrowein}. Moreover, oxygen has not been considered yet which is really important for yeast activity especially in the case of using Saccharomyces cerevisiae.\\
In this paper, the aspects of an effective model representing the process of wine fermentation are studied.
In the next section, based on a model taken from the literature, a new model with a yeast death term is derived. 
It follows a comparison of simulation results for both models in Section 3. Conclusions are presented in Section 4.
\section{Model derivation}
In this section, a model taken from the literature is further developed to take into account the yeast dynamics.  
As a starting point, we consider the model proposed by David et al. \cite{modeltempnitode}, that is described by the following differential equations:
\begin{equation}
\begin{cases}
\begin{aligned}
&\dfrac{\partial{X}}{\partial{t}}=\mu_{max}(T)\dfrac{N}{K_N+N}X\\
&\dfrac{\partial{N}}{\partial{t}}=-k_1\mu_{max}(T)\dfrac{N}{K_N+N}X\\
&\dfrac{\partial{E}}{\partial{t}}=\beta_{max}(T)\dfrac{S}{K_S+S}\dfrac{K_E(T)}{K_E(T)+E}X\\
&\dfrac{\partial{S}}{\partial{t}}=-k_2\beta_{max}(T)\dfrac{S}{K_S+S}\dfrac{K_E(T)}{K_E(T)+E}X
\end{aligned}
\end{cases}
\label{lit}
\end{equation}
with
\begin{equation}
\begin{aligned}
&X_0=X(0),\hspace{0.5cm}E_0=0,\\
&N_0=N(0),\hspace{0.5cm}S_0=S(0),\\
\end{aligned}
\end{equation}
This model is formulated with Micha\"elis-Menten kinetics. Here, $\mu_{max}(T)$ and $\beta_{max}(T)$ are the maximum specific growth rates which depend on temperature $T$; $K_N$ and $K_S$ are saturation constants, and $K_E(T)$ shows the ethanol inhibition dependent on the temperature. 
The parameters $k_1$ and $k_2$ are the yield coefficients associated to nitrogen and sugar, respectively.\\
In this model, the growth of yeast concentration is only dependent on the nitrogen concentration. Sugar is converted into ethanol and inhibited by it as well.\\
As this model does not represent all the phases of yeast cell population dynamics, we add a nonlinear term to the yeast concentration equation, 
thus ensuring that the lag and death phase of yeast cells take place. This behavior depends on the concentration of ethanol which inhibits the yeast such that if the ethanol concentration is below a tolerance $tol$ the number of yeast cells stays stationary, and if it is greater than $tol$, the yeast cells die. 
Furthermore, in the novel model, the sugar concentration is split up into the amount of sugar considered for the conversion into ethanol and the amount of sugar needed as a nutrient for the yeast.
Besides this, the presence of oxygen is added in the model because it is of paramount importance for yeast activity.\\
\begin{equation}
\begin{cases}
\begin{aligned}
\dfrac{\partial X}{\partial t}=&\mu_{max}(T) \dfrac{{N}}{K_N+N}\dfrac{S}{K_{S_1}+S}\\&\dfrac{O_2}{K_{O}+O_2}{X}-\Phi({E}){X}\\
\dfrac{\partial {N}}{\partial t}=&-k_1\mu_{max}(T) \dfrac{{N}}{K_N+{N}}\dfrac{{S}}{K_{S_1}+{S}}\\&\dfrac{{O_2}}{K_{O}+{O_2}}{X}\\
 \dfrac{\partial {E}}{\partial t}=&\beta_{max}(T) \dfrac{{S}}{K_{S_2}+{S}}\dfrac{K_E(T)}{K_E(T)+{E}}{X}\\
\dfrac{\partial{S}}{\partial t}=&-k_2\dfrac{\partial {E}}{\partial t}-k_3\mu_{max}(T) \dfrac{{N}}{K_N+N}\\&\dfrac{S}{K_{S_1}+S}\dfrac{O_2}{K_{O}+O_2}{X}\\
\dfrac{\partial {O_2}}{\partial t}=&-k_4\mu_{max}(T) \dfrac{{N}}{K_N+{N}}\dfrac{{S}}{K_{S_1}+{S}}\\&\dfrac{{O_2}}{K_{O}+{O_2}}{X}\\
\end{aligned}
\end{cases}
\label{new}
\end{equation}
with
\begin{equation}
\begin{aligned}
&X_0=X(0),\hspace{0.5cm}E_0=0,\\
&N_0=N(0),\hspace{0.5cm}S_0=S(0),\\
&O_{2_0}=O_2(0)
\end{aligned}
\end{equation}
In addition to the parameters already introduced for the starting model, the following parameters are needed. Instead of $K_S$ in this model two saturation constants associated to sugar, namely $K_{S_1}$ and $K_{S_2}$, are needed. Thereby $K_{S_1}$ represents the saturation constant associated to the part of sugar used as a nutrient for the yeast and $K_{S_2}$ is the saturation constant associated to the part of sugar needed for the metabolization into alcohol. Moreover, $k_3$ stands for the yield coefficient related to the part of sugar which is used as a nutrient for the yeast. $K_O$ is the Micha\"elis-Menten half-saturation constant associated to oxygen and $k_4$ represents the yield coefficient associated to oxygen.\\%
For $\Phi(E)$ the following term is considered:
\begin{equation}
\begin{aligned}
\Phi(E)=&\left(0.5 +\dfrac{1}{\pi}\arctan(k_{d1}(E-tol))\right)\\& k_{d2}(E-tol)^2
\end{aligned}
\label{phi1}
\end{equation}
where $tol$ is the tolerance of the ethanol concentration, and $k_{d1}$ and $k_{d2}$ are parameters associated to the death of yeast cells due to ethanol exceeding the tolerance $tol$.

\section{Results}
In the following, simulation results for the model by David et al. and our new model are compared.\\ 
The simulation was performed using ACADO toolkit - a toolkit for Automatic Control and Dynamic Optimization developed by Moritz Diehl et al. \cite{acadoMatlabManual, acadoManual}. In this context a BDF (Backward Differentiation Formula) integrator was used to solve the system of differential equations representing the process.\\
For the numerical results obtained, the parameters were chosen according to the following table.\\
\begin{table}[H]
\begin{tabular}{|l|l|l|}\hline
& (\ref{lit}) & (\ref{new}) with (\ref{phi1}) \\\hline
$\mu_1$ & $0.0126$ & $0.02$\\
$\mu_2$ & $0.0057$ & $0$ \\
$K_N$ & $0.0151$ & $0.0007$ \\
$k_1$ & $0.0513$ & $0.0115$ \\
$K_S$ & $7.6373$ & \\
$K_{S_1}$ & &$53.2669$\\
$K_{S_2}$ & &$0.1599$\\
$K_{E_1}$ & $0.0672$ & $1.5693$ \\
$K_{E_2}$ & $39.7925$ & $45.1692$ \\
$\beta_1$ & $0.2022$ & $0.1$ \\
$\beta_2$ & $0$ & $0.0728$ \\
$k_2$ & $2.1544$ & $1.9569$ \\ 
$k_{d_1}$ & & $9.9676$ \\
$k_{d_2}$ & & $0.0004$ \\
$k_3$ & & $2.8424$\\
$K_O$ & & $0.0002$\\
$k_4$ & & $0.0004$\\
$tol$ & &$70$\\
\hline
\end{tabular}
\caption{Parameter values for the different models}
\end{table}
In this series of simulation, the fermentation temperature is equal to $15^{\circ}{\rm C}$ for the first half-time of the process and $18^{\circ}{\rm C}$ for the second half-time of the process. Thereby $\mu_{max}(T)$, $\beta_{max}(T)$ and $K_E(T)$ are assumed to be linear dependent on temperature such that
\begin{equation}
\begin{aligned}
&\mu_{max}(T)=\mu_1T-\mu_2,\\ &\beta_{max}(T)=\beta_1T-\beta_2 \\&\text{ and}\\ & K_E(T)=-K_{E_1}T+K_{E_2}.
\end{aligned}
\end{equation}
For the differential states the following initial values were chosen.
\begin{table}[H]
\begin{tabular}{|l|l|}\hline
$X(0)$ & $0.2 g/l(\approx 400000/ml)$\\
$N(0)$ & $0.17 g/l$\\
$E(0)$ & $0 g/l$\\
$S(0)$ & $213.4 g/l$\\
$O_2(0)$ & $0.005 g/l$\\
\hline
\end{tabular}
\caption{Initial values for differential states}
\end{table}
\begin{figure}[H]
\includegraphics[scale=0.55]{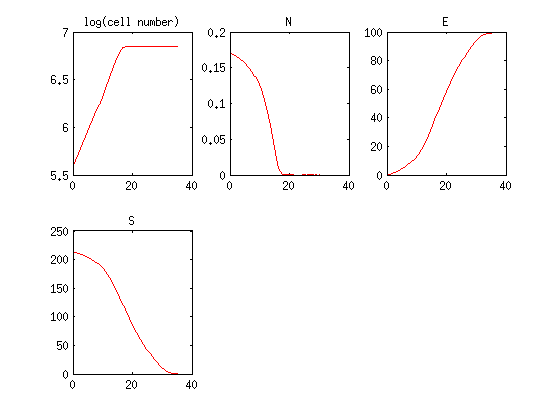}
\caption{Model by David et al. \cite{modeltempnitode}: Change of substrates and product relative to time}
\end{figure}
The model by David et al. \cite{modeltempnitode} represents the growth and lag phase observed in real processes for yeast behavior. Moreover, it is apparent that yeast growth just depends on the current concentration of assimilable nitrogen. This yields a stationary number of yeast cells at the point where the whole nitrogen is consumed. Alcohol and sugar stand in a certain relationship to each other, which is defined by the yield coefficient parameter $k_2$.\\
\begin{figure}[H]
\includegraphics[scale=0.55]{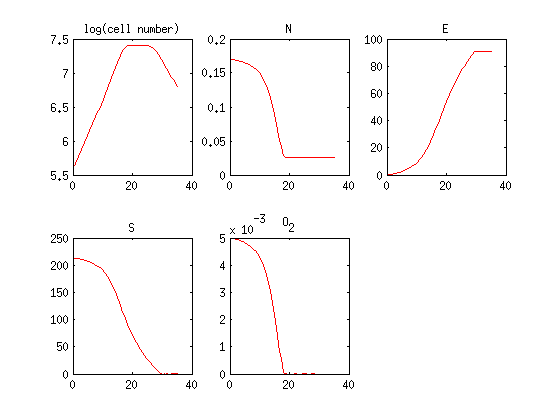}
\caption{New model with arctangent term (\ref{phi1}): Change of substrates and product relative to time}
\end{figure}
In addition to the latter model, the novel model with the death term (\ref{phi1}) includes the death of yeast cells where the concentration of alcohol is greater than the tolerance $tol$. This leads to results that represent the yeast behavior as it is observed in real experiments. The only phase missing is the first one which is not covered by this model.
Moreover, here, alcohol and sugar, except of what is needed as a nutrient for yeast, stand in a certain relationship to each other, which is defined by the yield coefficient parameter $k_2$.\\
\section{Conclusions}
In this paper, a new dynamical model for wine fermentation was derived. It was shown that the new model is able to predict 
the behavior of yeast growth as it is observed in real processes. Besides this, it also takes oxygen into account.

\section{Acknowledgements}
This research work was supported by the BMBF (German Federal Ministry of Education and Research) within the collaborative project R\OE NOBIO. \\
We would like to thank Peter Fürst, Peter Petter, Ivo Muha and Rainer Keicher for many useful comments. 
\bibliographystyle{plain}

\end{document}